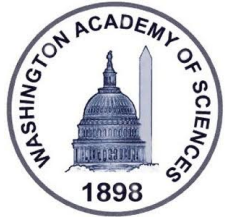

# HYDROGEN LINE OBSERVATIONS OF COMETARY SPECTRA AT 1420 MHZ

ANTONIO PARIS
THE CENTER FOR PLANETARY SCIENCE

## ABSTRACT

In 2016, the Center for Planetary Science proposed a hypothesis arguing a comet and/or its hydrogen cloud were a strong candidate for the source of the "Wow!" Signal. From 27 November 2016 to 24 February 2017, the Center for Planetary Science conducted 200 observations in the radio spectrum to validate the hypothesis. The investigation discovered that comet 266/P Christensen emitted a radio signal at 1420.25 MHz. All radio emissions detected were within 1° (60 arcminutes) of the known celestial coordinates of the comet as it transited the neighborhood of the "Wow!" Signal. During observations of the comet, a series of experiments determined that known celestial sources at 1420 MHz (i.e., pulsars and/or active galactic nuclei) were not within 15° of comet 266/P Christensen. To dismiss the source of the signal as emission from comet 266/P Christensen, the position of the 10-meter radio telescope was moved 1° (60 arcminutes) away from comet 266/P Christensen. During this experiment, the 1420.25 MHz signal disappeared. When the radio telescope was repositioned back to comet 266/P Christensen, a radio signal at 1420.25 MHz reappeared. Furthermore, to determine if comets other than comet 266/P Christensen emit a radio signal at 1420 MHz, we observed three comets that were selected randomly from the JPL Small Bodies database: P/2013 EW90 (Tenagra), P/2016 J1-A (PANSTARRS), and 237P/LINEAR. During observations of these comets, we detected a radio signal at 1420 MHz. The results of this investigation, therefore, conclude that cometary spectra are detectable at 1420 MHz and, more importantly, that the 1977 "Wow!" Signal was a natural phenomenon from a Solar System body.

## Introduction

On 15 August 1977, the Ohio State University Radio Observatory detected a strong narrowband signal in the constellation Sagittarius (Sgr) [1]. The frequency of the signal, which matched closely with the hydrogen line (1420.40575177 MHz), peaked at approximately 23:16:01 EDT [2]. On the same date and time, comet 266P/Christensen was transiting in the vicinity where the "Wow!" Signal was detected [3]. The purpose of this investigation, therefore, was to collect and analyze radio emission spectra and determine if comet 266P/Christensen and/or any other previously unknown celestial body in the Solar System was the source of the 1977 "Wow!" Signal. This investigation, moreover, was designed to improve our understanding of the content and origin of the "Wow!' Signal by determining if a neutral hydrogen cloud emitted from a short-period comet could be detected by a terrestrial radio telescope.

### The 10-Meter Radio Telescope and Spectrometer

For this experiment, we used a 10-meter radio telescope equipped with a spectrometer and a custom feed horn designed to collect a signal centered at 1420.25 MHz with a total bandwidth of 6.5 MHz. The radio telescope is composed of a frontend unit, which includes a low-noise preamplifier and a cylindrical feed horn for 1420 MHz. The 1420 MHz signal from the low-noise amplifier enters the rear panel (receiver backend) and is fed to a 1420 to 70 MHz dual conversion (internal) down converter. This converter has approximately 3 dB = 8 MHz bandwidth with the hydrogen rest frequency at 70.0 MHz. This 3 dB = 8 MHz wide IF signal is passed through a programmable gain IF amplifier and then split between the continuum square law detector and the spectrometer third conversion mixer. The programmable gain IF amplifier is used to compensate for feedline losses and to place the signal in the optimum range for the square law detectors [4].

The SpectraCyber software provides a control and hardware setting interface to the spectrometer. In addition, the software includes features which allow the observer to select between spectral and continuum observations, select time coordinate systems for the data files, set the computer clock to within 500 milliseconds, and change graphical features such as background or the amplitude axis on the data plot [4]. Final modifications to the software included a rapid scan rate mode, frequency shifting from 1420.406 MHz to a 70 MHz IF, and km/s displayed on the *x*-axis.

The graphs highlighted hereafter are "signal-averaged-intensity in volts (later converted to dB) *versus* time obtained directly from the 10-meter radio telescope with custom acquisition times, which varied depending on the observation [8]. To convert voltage changes to decibels and determine the strength of the signal's gain, the following logarithmic formula was used:

$$\text{dB} = 20\log (V_1/V_2)$$

The data collected was saved using the spreadsheet output format option of the SpectraCyber software and imported into Microsoft Excel as a text file. The data were then replotted and interpreted using the Chart Wizard feature in Microsoft Excel and converted into JPEG format.

**Testing the Hypothesis**

From 27 November 2016 to 24 February 2017, a series of observations in the neighborhood of the "Wow!" Signal coordinates was conducted to ascertain if the alleged "Wow!" Signal was a natural phenomenon rather than a signal from a source of extraterrestrial intelligence. These observations were made in both spectral and continuum modes (Table 1). In continuum mode, neutral hydrogen at rest with respect to the radio telescope's acquisition time appears as a peak along the *x*-axis while its intensity, in decibels (dB), appears along the *y*-axis. In the spectral mode, the neutral hydrogen line peaks that appear higher in frequency on the left side of the centered dashed line are due to neutral hydrogen gas receding (redshifted). Conversely, the neutral hydrogen line peaks that appear at frequencies higher on the right side of the centered dashed line are due to neutral hydrogen gas approaching (blueshifted).

Preliminary experiments from 27 November 2016 to 01 January 2017 were conducted to test, refine, and prepare for the observation of comet 266/P Christensen. The comet was then observed as it transited the area of the "Wow!" Signal between 20 January 2017 and 28 January 2017. In total, 200 observations (Table 1) were completed, which comprised of:

- The Galactic Plane near the coordinates of the 1977 "Wow!" Signal
- Radio Galaxy Cygnus A
- The Sun
- Comet 266/P Christensen as it transited near the coordinates of the "Wow!" signal
- Comet 266/P Christensen as it transited outside the coordinates of the "Wow!" signal.

- Clear Sky Observations near the coordinates of the "Wow!" Signal, and
- A baseline of random comets to determine if they emit a radio signal at 1420 MHz:

    o Comet P/2013 EW90 (Tenagra)
    o Comet P/2016 J1-A (PANSTARRS)
    o Comet 237P/LINEAR

**Table 1:**

List of observations conducted from 27 November 2016 to 24 February 2017.

| Date | Target | Type of Observation | Observations | 1420 MHz Detected |
|---|---|---|---|---|
| 27-Nov-16 | 1420 MHz Transmitter | Baseline Drift Scan to Test Feedhorn | 2 | Yes |
| 12-Dec-16 | Sun | Baseline Drift Scan to Test Feedhorn | 4 | Yes |
| 14-Dec-16 | Cygnus A | Baseline Drift Scan to Test Feedhorn | 5 | Yes |
| 5-Jan-17 | Sun | Baseline Drift Scan to Test Feedhorn | 4 | Yes |
| 9-Jan-17 | Cygnus A | Baseline Drift Scan | 5 | Yes |
| 10-Jan-17 | Cygnus A | Baseline Drift Scan | 5 | Yes |
| 11-Jan-17 | Galactic Plane | Baseline Drift Scan | 5 | Yes |
| 12-Jan-17 | Galactic Plane | Baseline Drift Scan | 5 | Yes |
| 12-Jan-17 | Galactic Plane | Baseline Drift Scan | 5 | Yes |
| 13-Jan-17 | Sun | Baseline Drift Scan | 4 | Yes |
| 13-Jan-17 | 266/P | Drift Scan | 4 | Yes |
| 14-Jan-17 | Galactic Plane | Baseline Drift Scan | 5 | Yes |
| 14-Jan-17 | 266/P | Drift Scan | 4 | Yes |
| 19-Jan-17 | Sun | Baseline Drift Scan | 4 | Yes |
| 19-Jan-17 | Galactic Plane | Baseline Drift Scan | 5 | Yes |
| 19-Jan-17 | 266/P | Drift Scan | 4 | Yes |
| 20-Jan-17 | Sun | Baseline Drift Scan | 4 | Yes |
| 20-Jan-17 | Galactic Plane | Baseline Drift Scan | 5 | Yes |
| 20-Jan-17 | 266/P (at Wow! Signal Coordinates) | Drift Scan | 5 | Yes |
| 26-Jan-17 | Sun | Baseline Drift Scan | 4 | Yes |
| 26-Jan-17 | Cygnus A | Baseline Drift Scan | 5 | Yes |
| 26-Jan-17 | 266/P (at Wow! Signal Coordinates) | Drift Scan | 4 | Yes |
| 27-Jan-17 | Sun | Baseline Drift Scan | 5 | Yes |
| 27-Jan-17 | Cygnus A | Baseline Drift Scan | 4 | Yes |
| 27-Jan-17 | 266/P (at Wow! Signal Coordinates) | Drift Scan | 4 | Yes |
| 27-Jan-17 | Clear Sky | Baseline Drift Scan | 1 | No |
| 28-Jan-17 | Sun | Baseline Drift Scan | 2 | Yes |
| 28-Jan-17 | Clear Sky (Wow! Signal Coordinates) | Baseline Drift Scan | 4 | No |
| 28-Jan-17 | 266/P (at Wow! Signal Coordinates) | Drift Scan | 5 | Yes |
| 28-Jan-17 | Sun | Baseline Drift Scan | 2 | Yes |
| 28-Jan-17 | Galactic Plane | Baseline Drift Scan | 5 | Yes |
| 29-Jan-17 | Clear Sky (Wow! Signal Coordinates) | Baseline Drift Scan | 5 | No |
| 30-Jan-17 | Clear Sky (Wow! Signal Coordinates) | Baseline Drift Scan | 5 | No |
| 30-Jan-17 | Clear Sky (Wow! Signal Coordinates) | Baseline Drift Scan | 5 | No |
| 30-Jan-17 | Clear Sky (Wow! Signal Coordinates) | Baseline Drift Scan | 5 | No |
| 22-Feb-17 | Comet P/2013 EW90 | Baseline Drift Scan | 15 | Yes |
| 23-Feb-17 | Comet P/2016 J1-A | Baseline Drift Scan | 16 | Yes |
| 24-Feb-17 | Comet 237P/LINEAR | Baseline Drift Scan | 20 | Yes |
| | | **Total Observations** | **200** | |

During all phases of this investigation, the 10-meter radio telescope was tested and calibrated daily to ensure accuracy. In addition, the data collected were handwritten and logged into the Center for Planetary Science Observation log, which was then typed and saved as a PDF file. As a reference for the reader, all altitudes and azimuths noted hereafter are in relation to the Site-B Observatory, which is in Wesley Chapel, FL, and the coordinates for comets 266/P Christensen, P/2013 EW90 (Tenagra), P/2016 J1-A (PANSTARRS), and 237P/LINEAR were obtained from the JPL Small Bodies Database.

## Baseline Data Collection of Radio Galaxy Cygnus

To establish a baseline for 1420 MHz, and to disqualify potential noise, the 10-meter radio telescope was directed toward Cygnus A during several observing days (Table 1). The galaxy Cygnus A, which is 263 megaparsecs away [5], is the strongest known celestial radio source, other than the Sun, in the sky. Cygnus A contains an active galactic nucleus. The source of this strong radio signal is the dense region at the center of the galaxy, which has a much greater than normal luminosity of the electromagnetic spectrum [6]. This extreme emission, which is thought to be a result of the accretion of matter by a supermassive black hole [6], can be observed at 1420 MHz. The Site-B 10-meter radio telescope was directed toward Cygnus A on five separate dates for a total of 24 observations (Table 1). The purpose of these observations was to determine the accuracy of the radio telescope's position toward known celestial coordinates, to acquire a baseline reading at 1420 MHz, and to collect data from Cygnus A in both spectral and continuum mode.

## Cygnus A Profile in Spectral Mode

All observations of Cygnus A (Table 2) produced strong radio emissions at 1420 MHz. On average, the baseline signal as Cygnus A drifted into the lobe of the telescope increased by 9.70 dB (Figure 1). As Cygnus A drifted away from the lobe of the telescope, the signal decreased by 16.4 dB. The observations of Cygnus A, therefore, concluded that the 10-meter radio telescope's azimuth and altitude indicators were optimally functional, the 10-meter radio telescope detected radio emissions at 1420 MHz, and a spectral baseline of neutral hydrogen from a known celestial source could be later compared with comet 266/P Christensen.

**Table 2**: Celestial Coordinates of Galaxy Cygnus A on 14 December 2016 at 1222 EST

| | |
|---|---|
| Right Ascension | $20^h\,00^m\,03.49^s$ |
| Declination | $+40°\,46'\,58.1"$ |
| Altitude | $+56°\,27'\,28.0"$ |
| Azimuth | $057°\,58'\,40.6"$ |
| Galactic Longitude | $076°\,11'\,23.3"$ |
| Galactic Latitude | $+05°\,45'\,19.6"$ |

**Figure 1:** 1420 MHz Radio Emission from Galaxy Cygnus A on 14 December 2016 at 1222 EST

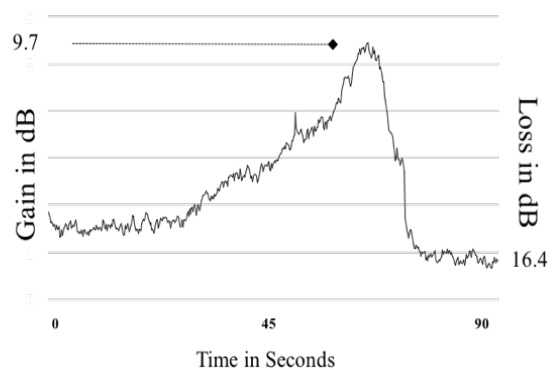

## Cygnus A Profile in Continuum Mode

Observations of Cygnus A (Table 3) in continuum mode produced strong radio emissions at 1420 MHz. On average, the baseline signal as Cygnus A drifted into the lobe of the telescope increased by 10.20 dB (Figure 2). As Cygnus A drifted away from the lobe of the telescope, the signal decreased by 9.11 dB. The observations of Cygnus A, therefore, demonstrated that the 10-meter radio telescope's azimuth and altitude indicators were optimally functional, the 10-meter radio telescope detected radio emissions at 1420 MHz, and a continuum baseline of neutral hydrogen from a known celestial source could be later compared with comet 266/P Christensen.

**Table 3:** Celestial Coordinates of Radio Galaxy Cygnus A on 10 January 2017 at 1416 EST

**Figure 2:** Radio Galaxy Cygnus A on 10 January 2017 at 1416 EST

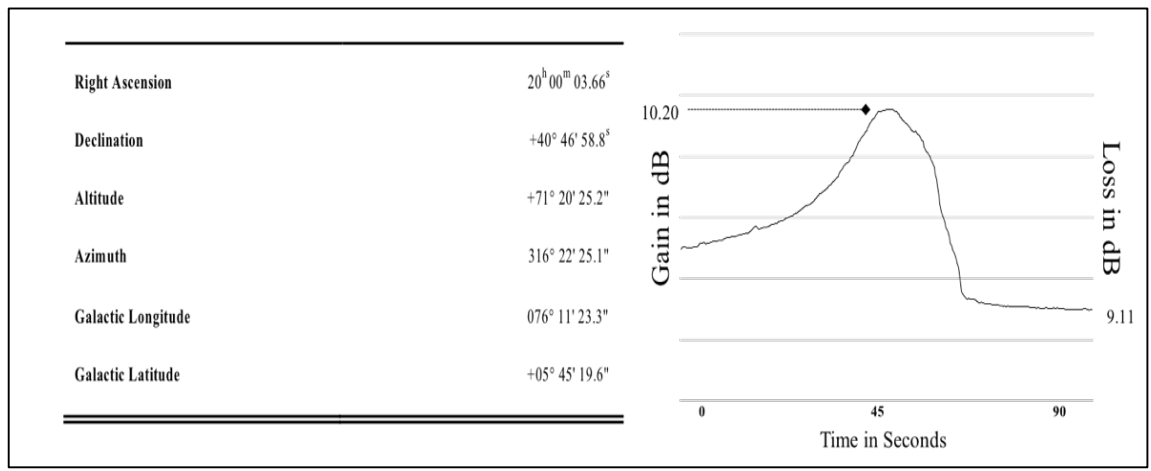

## Baseline Data Collection of the Galactic Plane

The Site-B 10-meter radio telescope was directed toward the Milky Way's Galactic Plane (Table 4) on six separate dates for a total of 30 observations (Figure 3). The Milky Way's Galactic Plane is the plane in which most of the galaxy's mass lies. The Galactic Plane is the most common known source of strong radio emissions in the electromagnetic spectrum, including at 1420 MHz [7]. Throughout this experiment, the edge of the Galactic Plane, as viewed from an East to West azimuth, was 23° from the coordinates of the "Wow!" Signal and the location of comet 266/P Christensen.

The purpose of these observations was to determine the accuracy of the radio telescope's position toward the sky, to conduct a baseline reading at 1420 MHz, and to locate and identify known/unknown celestial radio sources that could account for being the source of the 1977

"Wow!" Signal. The acquisition times for these sets of observations were 40 minutes, which provided sufficient time for the Galactic Plane to drift completely over the lobe of the telescope.

**Table 4:**

Drift Scan Celestial Coordinates for Galactic Plane Observations

|        | RA Start      | Declination Start | RA End        | Declination End |
|--------|---------------|-------------------|---------------|-----------------|
| **Scan A** | 17h58m03.77s | –18°58'54.7"      | 18h57m43.31s  | –17°25'06.4"    |
| **Scan B** | 17h50m59.16s | –24°12'30.6"      | 18h55m24.84s  | –24°44'39.3"    |
| **Scan C** | 17h48m37.55s | –27°49'59.5"      | 18h54m11.13s  | –27°43'56.0'    |

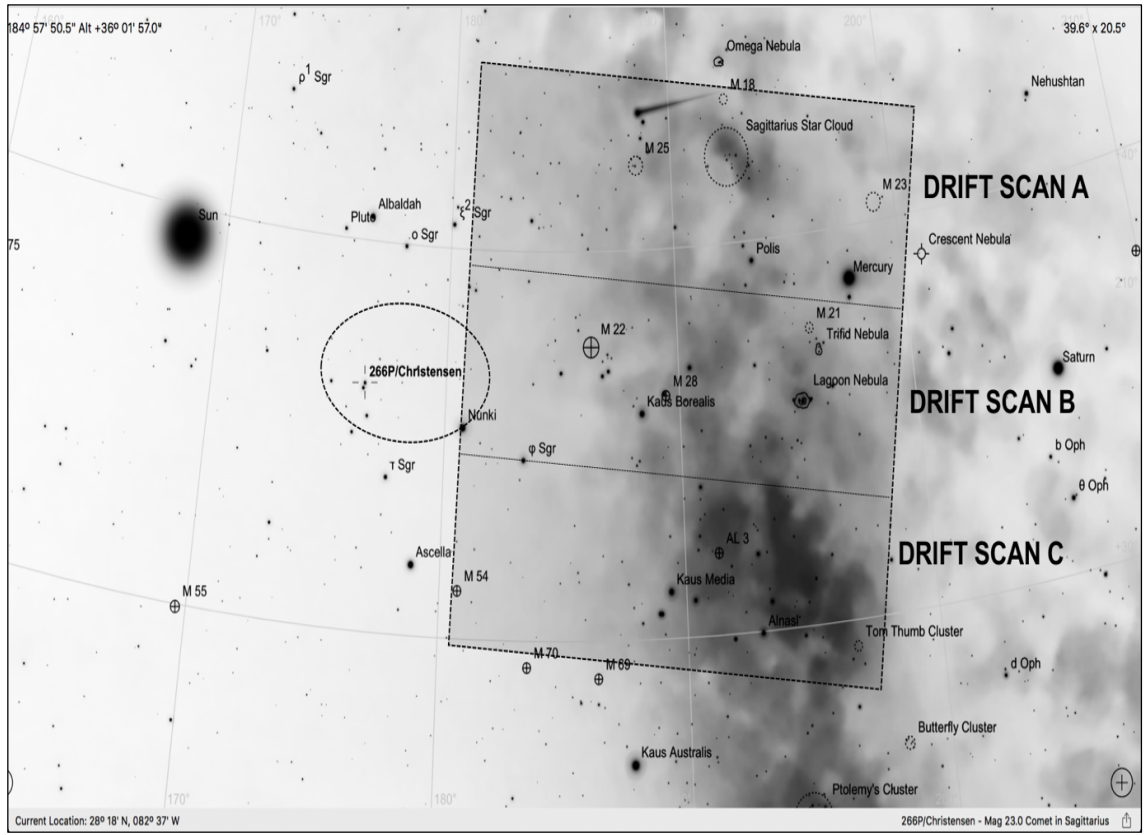

**Figure 3:** Galactic Plane Drift Scan Locations A, B, and C.

## Galactic Plane Profile in Spectral Mode

Observations of the Galactic Plane in areas A, B, and C (Figures 4, 5, and 6) did not detect strong radio sources at 1420 MHz. The gradual increase of the baseline signal was due to hydrogen gas approaching radially (blueshifted).

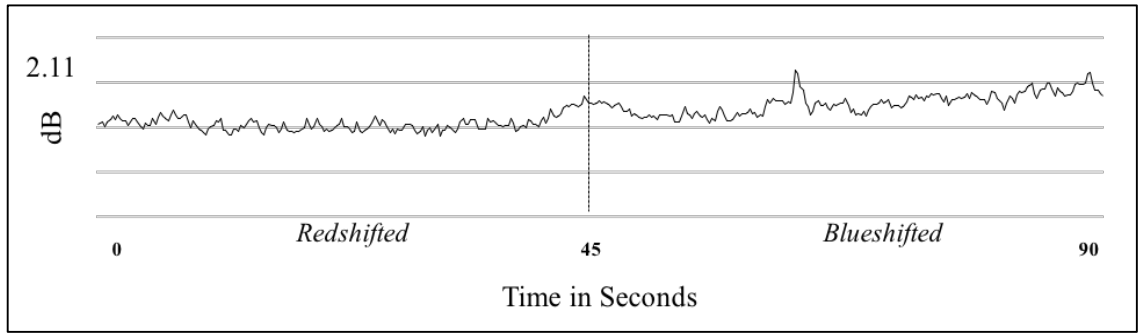

**Figure 4:** Spectral Drift "A" Scan of Galactic Plane on 12 January 2017

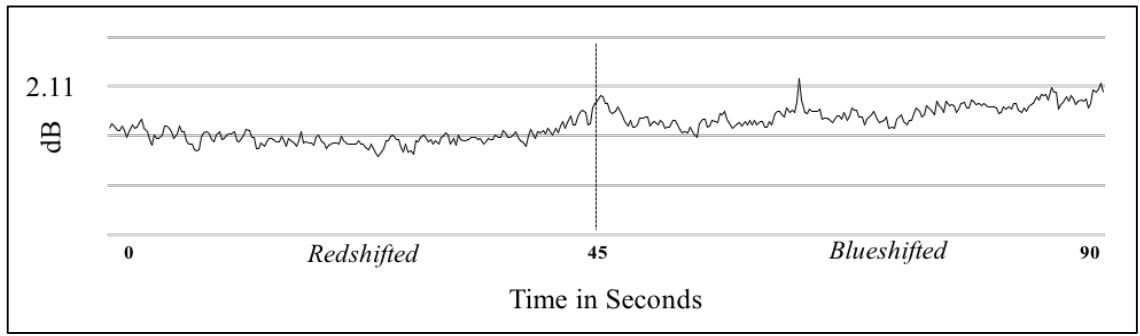

**Figure 5:** Spectral Drift "B" Scan of Galactic Plane on 19 January 2017

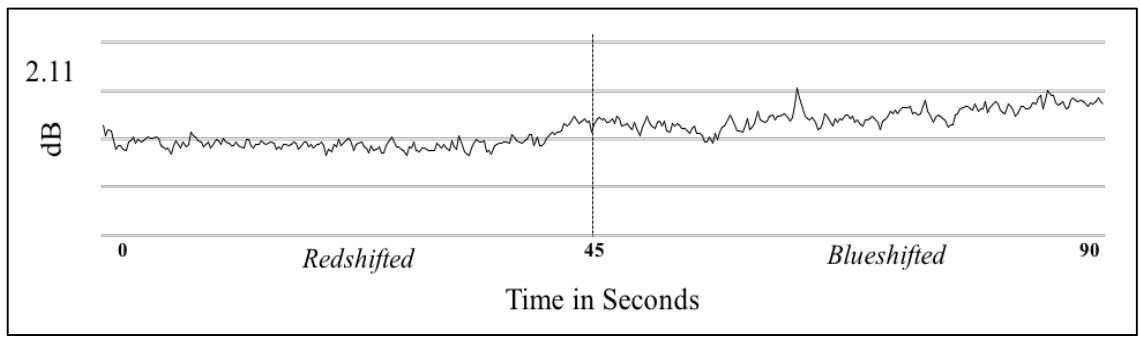

**Figure 6:** Spectral Drift "C" Scan of Galactic Plane on 20 January 2017

## Galactic Plane Profile in Continuum Mode

Observations of the Galactic Plane in areas A, B, and C detected a strong radio source at 1420 MHz (Figure 7). To acquire the Galactic Plane "edge on" as it drifted into and out of the lobe of the telescope, an acquisition time of 40 minutes was required. On average, the baseline signal as the Galactic Plane drifted into the lobe of the telescope increased by 6.36 dB (Figure 7). As the Galactic Plane drifted away from the lobe of the telescope, the signal decreased by 11.45 dB.

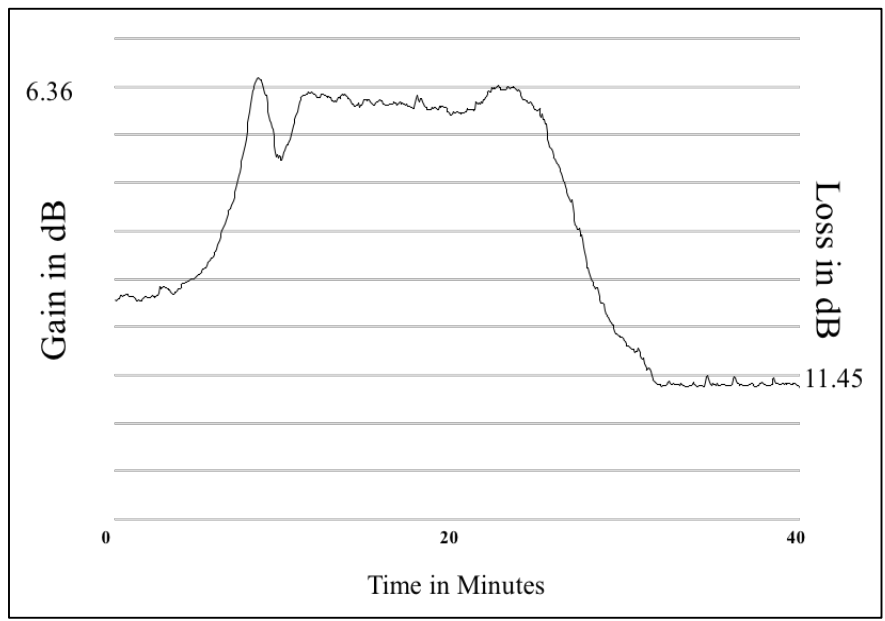

**Figure 7:** Drift Scan of Galactic Plane on 14 January 2017 at 1200 EST (Continuum Mode)

## Baseline Data Collection of the Sun

From 27 November 2016 to 30 January 2017, 33 observations of the Sun were completed (Table 5). The purpose of these observations was to obtain a baseline reading of neutral hydrogen and to confirm the accuracy of the telescope's altitude and azimuth indicators prior to every observation of comet 266/P Christensen. All observations of the Sun detected a strong neutral hydrogen signal. The strength of the neutral hydrogen is represented as a single line moving up along the *y*- and *x*-axes (Figure 8) as the Sun drifted into the lobe of the telescope.

**Table 5:** Celestial Coordinates of the Sun 12 December 2016 at 1200 EST (Continuum Mode)

**Figure 8:** Drift Scan of the Sun on 12 December 2016 at 1200 EST (Continuum Mode)

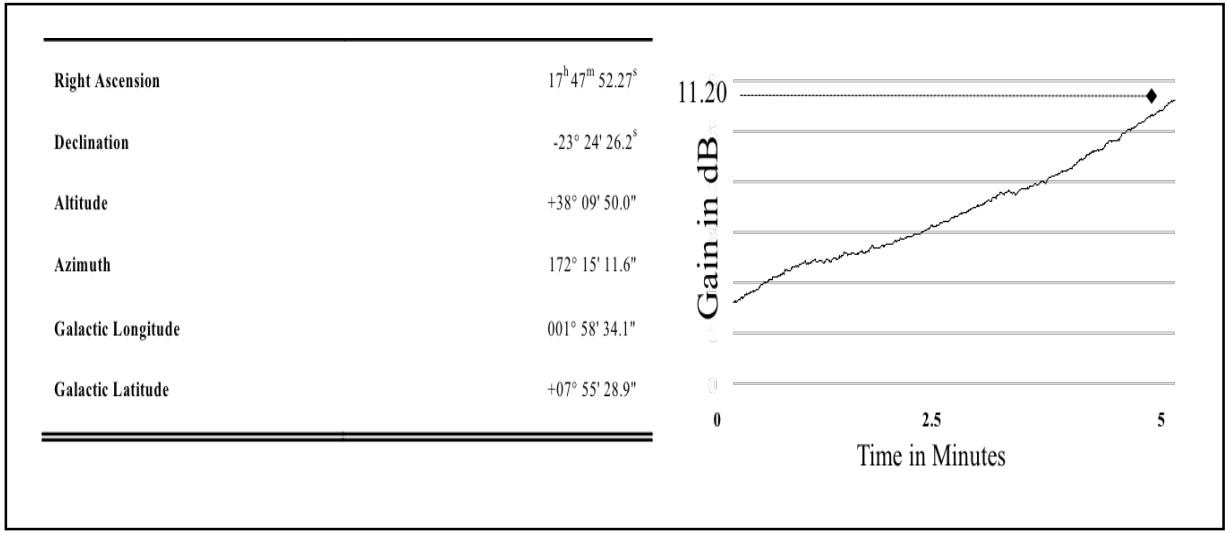

| | |
|---|---|
| Right Ascension | 17$^h$ 47$^m$ 52.27$^s$ |
| Declination | -23° 24' 26.2$^s$ |
| Altitude | +38° 09' 50.0" |
| Azimuth | 172° 15' 11.6" |
| Galactic Longitude | 001° 58' 34.1" |
| Galactic Latitude | +07° 55' 28.9" |

## Observations of Comet 266/P Christensen

From 20–28 January 2017, the 10-meter radio telescope was directed at comet 266/P Christensen (Figure 9) as it transited near the coordinates of the "Wow!" Signal. During all observations (Tables 6, 7, and 8) a radio signal with varying intensity at 1420.25 MHz was detected. On 20 January 2017, at 1100 EST, as comet 266/P Christensen drifted into the lobe of the telescope, the baseline signal increased by 7.95 dB. As the comet drifted away from the lobe of the telescope, the signal decreased by 11.48 dB (Figure 10). On 26 January 2017, at 1100 EST, as comet 266/P Christensen drifted into the lobe of the telescope, the baseline signal increased by 8.32 dB. As the comet drifted away from the lobe of the telescope, the signal decreased by 7.80 dB (Figure 11). The final observation was conducted on 28 January 2017 at 1235 EST. As comet 266/P Christensen drifted into the lobe of the telescope, the baseline signal increased by 9.97 dB. As the comet drifted away from the lobe of the telescope, the signal decreased by 11.42 dB (Figure 12).

**Figure 9:**

Date of Observations of Comet 266/P Christensen in relation to the coordinates of the Wow! Signal in 1977

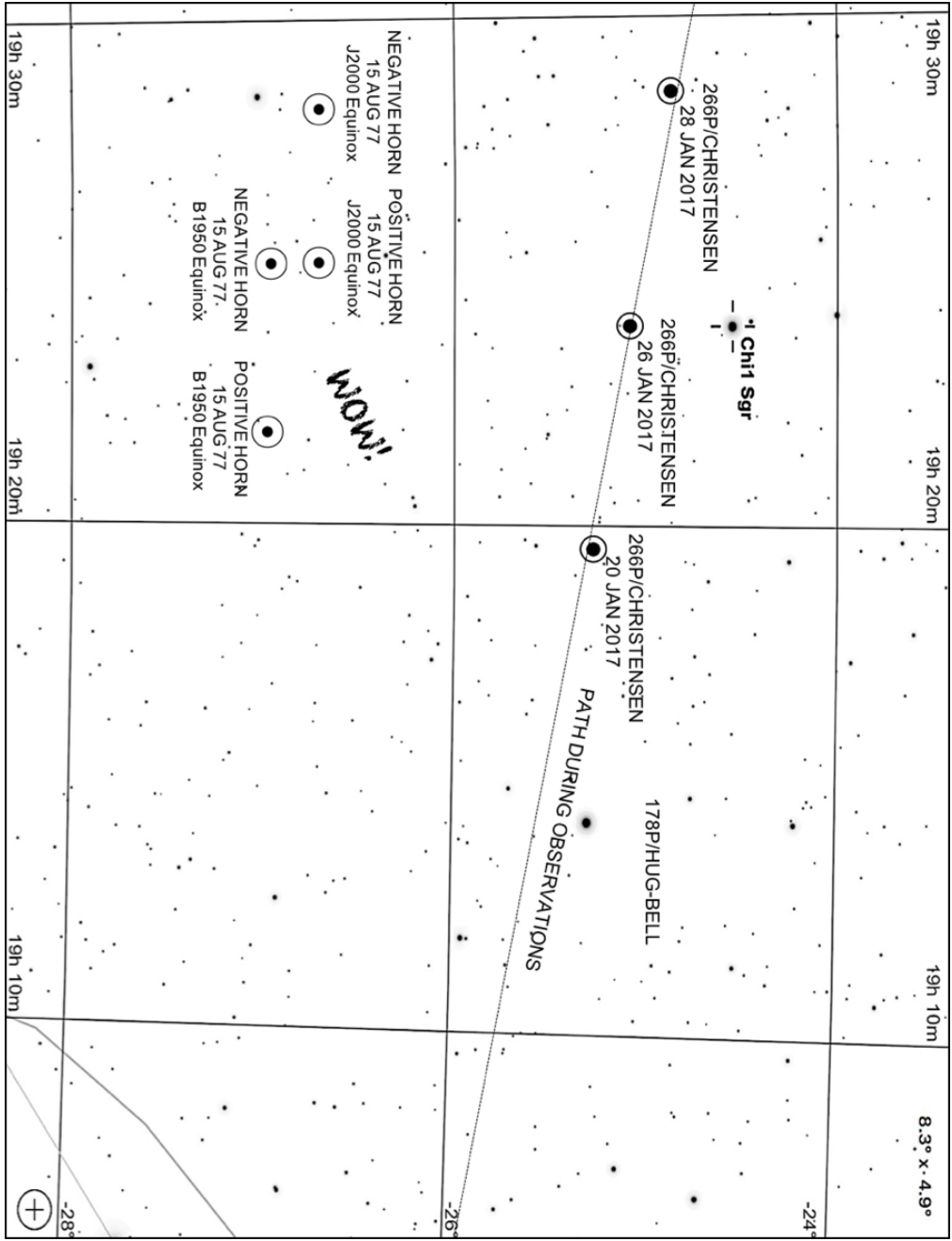

**Table 6:** Celestial Coordinates of Comet 266/P Christensen on 20 January 2017 at 1100 EST

| | |
|---|---|
| Right Ascension | $19^h 19^m 44.21^s$ |
| Declination | $-25° 00' 23.1^s$ |
| Altitude | $+35° 24' 53.3"$ |
| Azimuth | $166° 18' 20.1"$ |
| Galactic Longitude | $012° 55' 08.0"$ |
| Galactic Latitude | $-16° 46' 50.9"$ |

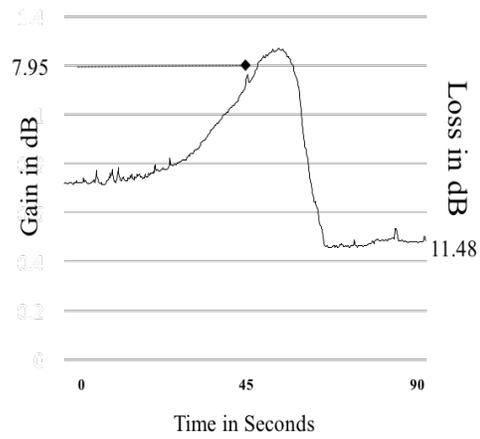

**Figure 10:** Drift Scan of Comet 266/P Christensen on 20 January 2017 at 1100 EST

**Table 7:** Celestial Coordinates of Comet 266/P Christensen on 26 January 2017 at 1100 EST

| | |
|---|---|
| Right Ascension | $19^h 25^m 54.97^s$ |
| Declination | $-24° 48' 54.9^s$ |
| Altitude | $+36° 21' 25.8"$ |
| Azimuth | $171° 03' 39.3"$ |
| Galactic Longitude | $013° 39' 41.9"$ |
| Galactic Latitude | $-18° 00' 14.6"$ |

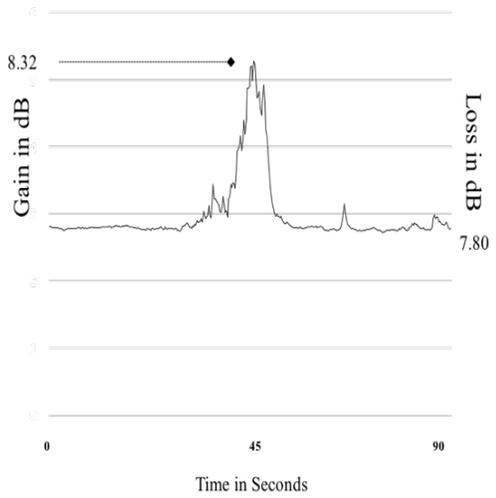

**Figure 11:** Drift Scan of Comet 266/P Christensen on 26 January 2017 at 1100 EST

**Table 8:** Celestial Coordinates of Comet 266/P Christensen on 28 January 2017 at 1235 EST

| | |
|---|---|
| Right Ascension | $19^h 28^m 01.32^s$ |
| Declination | $-24° 44' 51.4^s$ |
| Altitude | $+34° 24' 38.1"$ |
| Azimuth | $199° 08' 19.8"$ |
| Galactic Longitude | $013° 55' 59.2"$ |
| Galactic Latitude | $-18° 25' 17.9"$ |

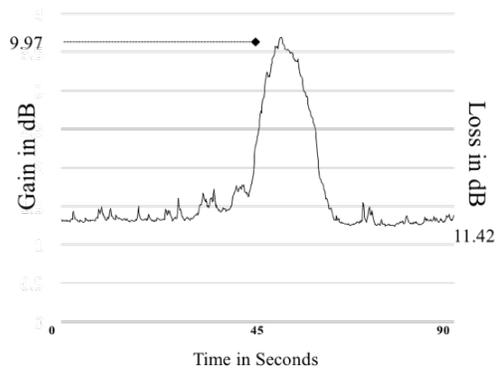

**Figure 12:** Drift Scan of Comet 266/P Christensen on 28 January 2017 at 1235 EST

**Repositioning of the Telescope During Observations of Comet 266/P Christensen**

On 28 January 2017, at 1130 EST and 1200 EST, the 10-meter radio telescope was directed at comet 266/P Christensen. During both observations, a radio signal at 1420.25 MHz was detected. To dismiss the radio signal as emanating from comet 266/P Christensen, the radio telescope was repositioned "away from" and "back to" the comet. In the first experiment, the radio telescope's azimuth was moved 1° (60 arcminutes) away from comet 266/P Christensen. As the telescope rotated away from the comet, the strength of the radio signal decreased by 10.12 dB (Figure 13). Conversely, when the telescope rotated 1° (60 arcminutes) back to the comet, the radio signal increased by 13.39 dB (Figure 14). The second set of experiments was centered on rotating the telescope's altitude 1° (60 arcminutes) away from comet 266/P Christensen. When the telescope rotated away from the comet the signal decreased by 9.3 dB. When the telescope was redirected 1° (60 arcminutes) back to the comet, the signal increased by 8.7 dB (Figure 15).

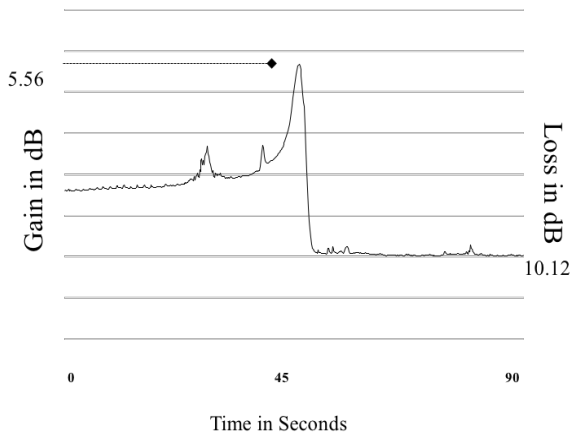

**Figure 13:** Repositioning of the Telescope's Azimuth 1° (60 arcminutes) away from Comet 266/P Christensen

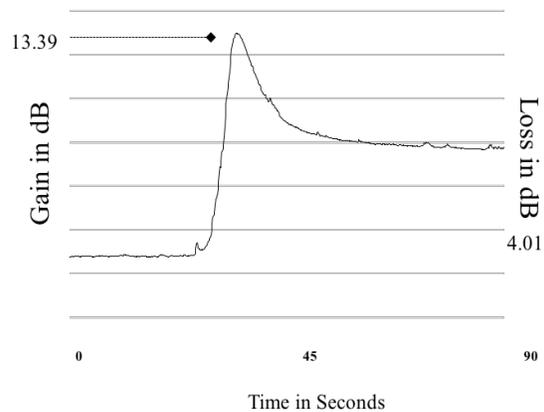

**Figure 14:** Repositioning of the Telescope's Azimuth 1° (60 arcminutes) back toward Comet 266/P Christensen

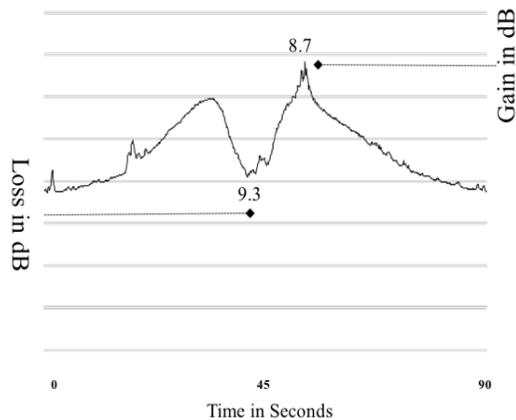

**Figure 16:** Repositioning of the Telescope's Altitude up and down consecutively 1° (60 arcminutes)

## Signal to Noise Ratio of Comet 266/P Christensen and the "Wow!" Signal

The maximum value of the "Wow!" Signal was 30 sigma (i.e., 30 times the standard deviation of the data). The peak of the "Wow!" Signal, moreover, was about 30.76 sigma. This also means that the signal-to-noise ratio was about 30.76 at the peak.

By using the signal to noise ratio (snr) formula (Figure 16), where pSignal is the strongest signal strength we detected from comet 266/P Christensen (5.21 volts on 20 January 2017), pNoise is the background noise detected during the same observation of comet 266/P Christensen (3.55, 3.44, 3.14, 3.98 V), μ is the mean and σ is the standard deviation (sigma), we calculated that the snr for comet 266/P Christensen was 4.76. We infer, therefore, that the signal we detected during observations of comet 266/P Christensen was not background noise. We speculate that the strength of the original "Wow!" Signal in 1977 (30 sigma) would have been accounted for by the size of the Big Ear Radio Telescope (when compared with Site B's 10-meter telescope) and/or the potential loss of mass from comet 266/P Christensen, which would have been considerably larger 40 years ago.

**Figure 16:** Signal to Noise Ratio Formula

$$snr = \frac{pSignal}{pNoise} = \frac{\mu}{\sigma} \quad \text{whereas} \quad \mu = \frac{x1+x2...+Xn}{n} \quad \text{and} \quad \sigma = \sqrt{\frac{1}{n-1} \sum_{i=1}^{n} (x_i-\mu)^2}$$

$$\mu = \frac{3.55 + 3.44 + 3.14 + 5.21 + 3.98}{5} = 3.864$$

$$\sigma = \sqrt{\frac{1}{n-1} \sum_{i=1}^{n} (x_i-\mu)^2} = \sum_{i=1}^{n} (x_i-\mu)^2 = (3.55 - \mu)^2 + (3.44 - \mu)^2 + (3.14 - \mu)^2 + (5.21 - \mu)^2 + (3.98 - \mu)^2$$

$$\sigma = \sum_{i=1}^{n} (x_i-\mu)^2 = 0.811$$

$$\frac{\mu}{\sigma} = \frac{3.864}{0.811} = 4.76$$

## Measurements of Clear Sky in Sagittarius

From 28–30 January 2017, 24 clear sky observations in continuum mode were conducted. The purpose of these observations was twofold: to determine if any radio signals at 1420 MHz could be detected after comet 266/P Christensen transited out of the area where the "Wow!" Signal was detected, and to acquire a baseline of background noise near the coordinates of the "Wow!" Signal. During these observations (Figures 17, 18, and 19), the baseline signal remained stable from 1.06 dB to 1.07 dB and no significant radio emission at 1420 MHz, other than background and system noise, was detected.

**Figure 17:**

Clear Sky Drift Scan on 28 January 2017 where comet 266/P Christensen was on 20 January 2017

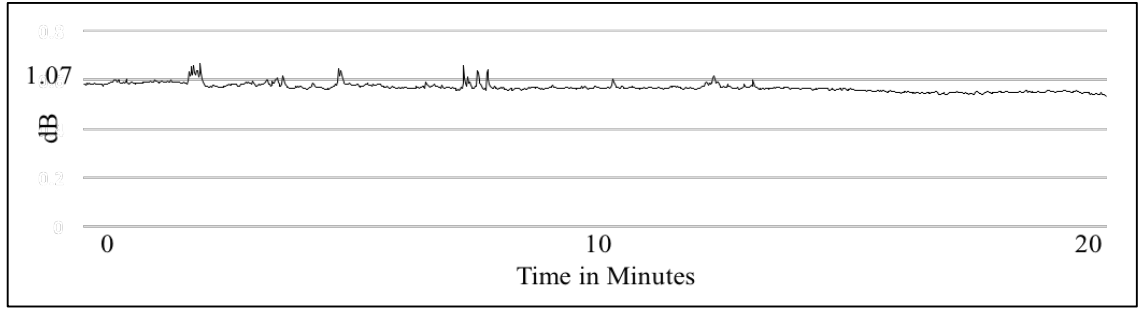

**Figure 18:**

Clear Sky Drift Scan on 29 January 2017 where comet 266/P Christensen was on 26 January 2017

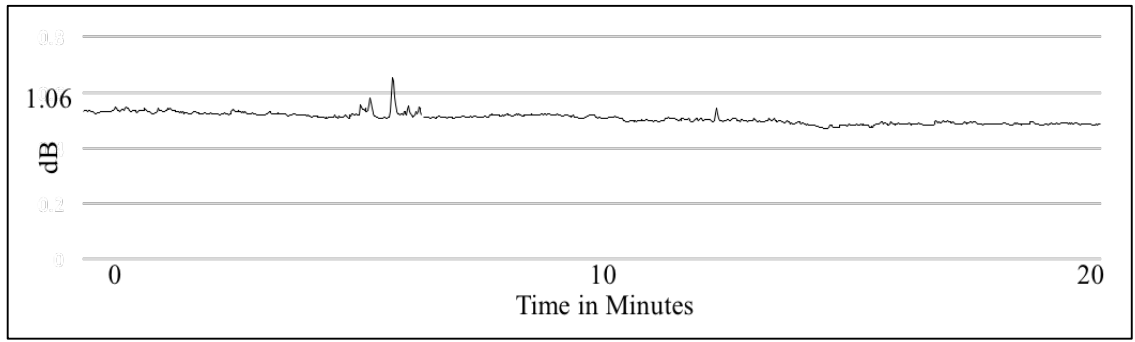

**Figure 19:**

Clear Sky Drift Scan on 30 January 2017 where comet 266/P Christensen was on 28 January 2017

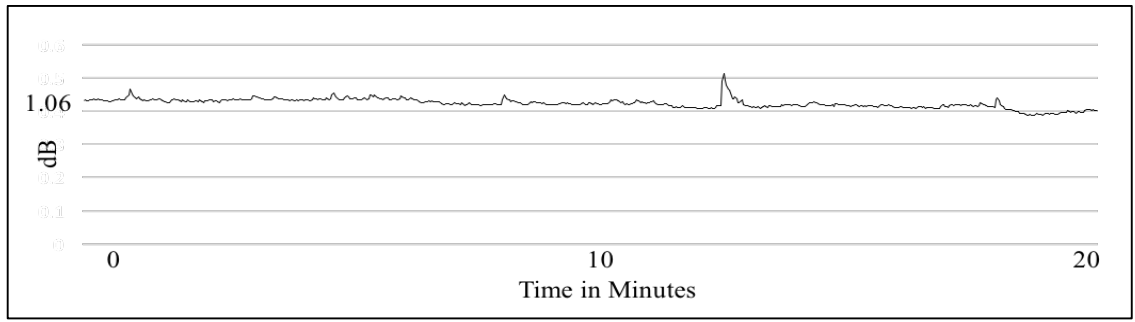

**Observations on Three Random Comets**

From 22–24 February 2017, the 10-meter radio telescope was directed at three random comets selected from the JPL Small Bodies Database. The comets selected were P/2013 EW90 (Tenagra), P/2016 J1-A (PANSTARRS), and 237P/LINEAR. The purpose of these observations was to determine if comets other than comet 266/P Christensen emitted a radio signal at 1420 MHz. During observations, these comets were more than 10° from the Sun and more than 20° from the Galactic Plane.

On 22 February 2017, the 10-meter radio telescope made 15 observations of comet P/2013 EW90 (Tenagra) as it transited the constellation Aquarius. During these observations, a radio signal with varying intensity at 1420 MHz was detected. At 1100 EST, as comet P/2013 EW90 (Tenagra) drifted into the lobe of the telescope, the baseline signal increased by 5.62 dB (Figure 20). As the comet drifted away from the lobe of the telescope, the signal decreased by 5.67 dB. All 15 observations of P/2013 EW90 (Tenagra) detected a radio signal at 1420 MHz.

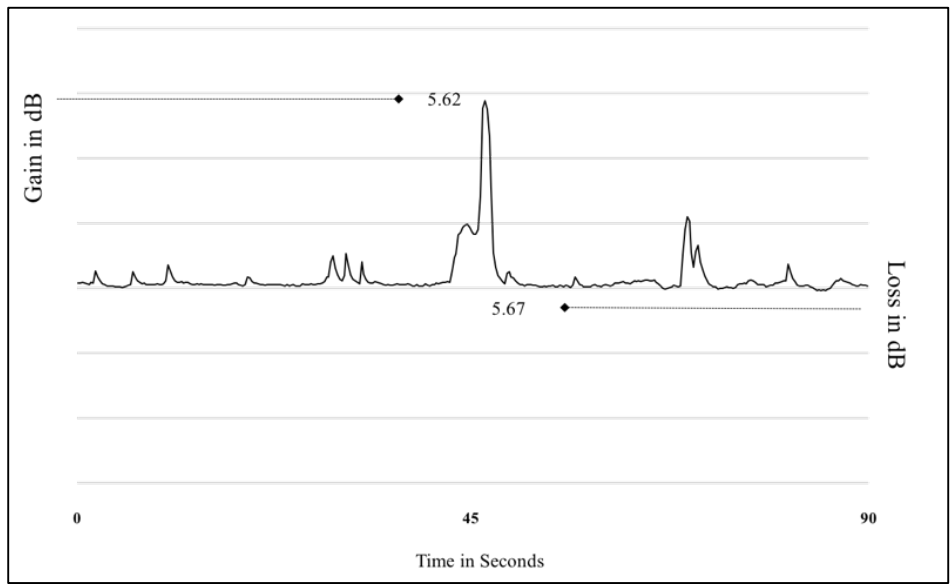

**Figure 20:**

Drift Scan of Comet P/2013 EW90 (Tenagra) on 22 February 2017 at 1100 EST

On 22 February 2017, the 10-meter radio telescope made 16 observations of comet P/2016 J1-A (PANSTARRS) as it transited the constellation Aquarius. During these observations, a radio signal with varying intensity at 1420 MHz was detected. At 1130 EST, as comet P/2016 J1-A (PANSTARRS) drifted into the lobe of the telescope, the baseline signal increased by 5.64 dB (Figure 21). As the comet drifted away from the lobe of the telescope, the signal decreased by 5.22 dB. All 16 observations of P/2016 J1-A (PANSTARRS) detected a radio signal at 1420 MHz.

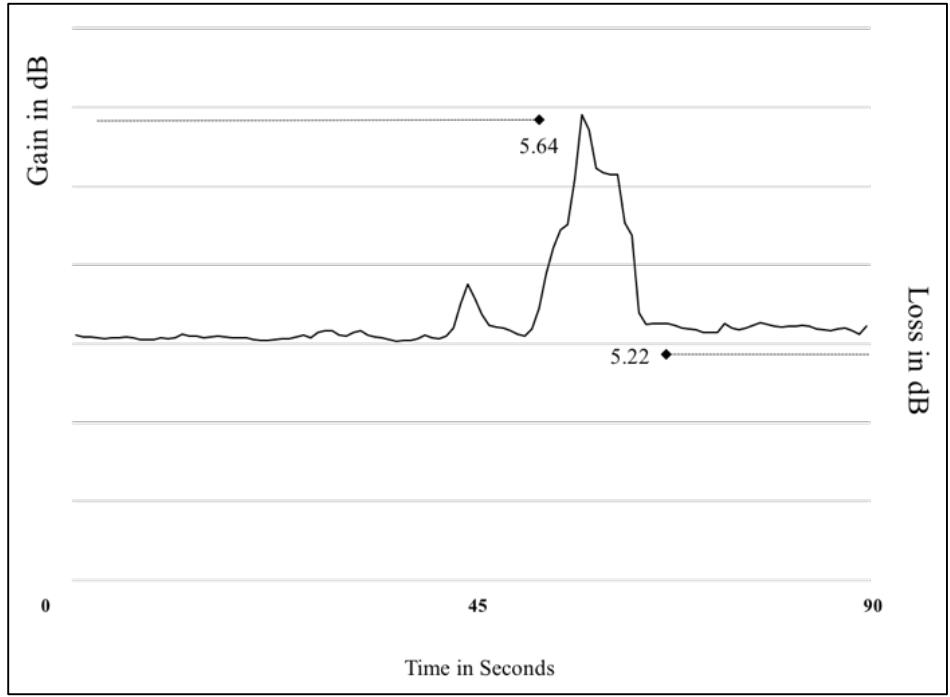

**Figure 21:**

Drift Scan of Comet P/2016 J1-A (PANSTARRS) on 23 February 2017 at 1130 EST

On 24 February 2017, the 10-meter radio telescope made 20 observations of comet 237P/LINEAR as it transited the constellation Aquarius. During these observations, a radio signal with varying intensity at 1420 MHz was detected. At 1230 EST, as comet 237P/LINEAR drifted into the lobe of the telescope, the baseline signal increased by 5.40 dB (Figure 22). As the comet drifted away from the lobe of the telescope, the signal decreased by 5.63 dB. All 20 observations of 237P/LINEAR detected a radio signal at 1420 MHz.

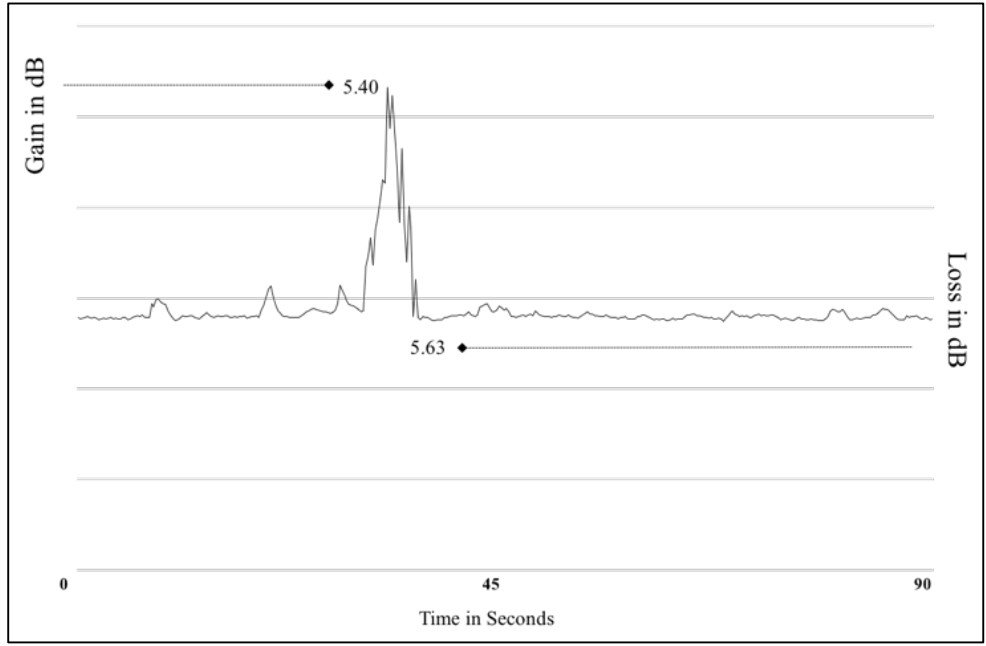

**Figure 22:**

Drift Scan of Comet 237P/LINEAR on 24 February 2017 at 1230 EST

## Analysis and Interpretation of Data

Upon completion of the 200 observations, a comparison of the neutral hydrogen lines from Cygnus A, the Galactic Plane, the Sun, comet 266/P Christensen, and the clear sky were conducted. An analysis (Figure 23) of these spectra established a well-defined distinction between the signals acquired during this investigation. Radio celestial sources such as Cygnus A, for example, illustrate a strong increase and decrease of dB as the radio signal drifted into and out of the lobe of the telescope. When compared with the spectra from comet 266/P, there is a clear (but weaker) similarity between the spectra of the comet and Cygnus A. Spectra of the Galactic Plane, the Sun, and clear skies, however, presented no similarities when compared with the comet. We conclude, therefore, that the neutral hydrogen signal acquired during observations of comet 266/P Christensen were not from a known celestial source and, more importantly, were emitted by the comet.

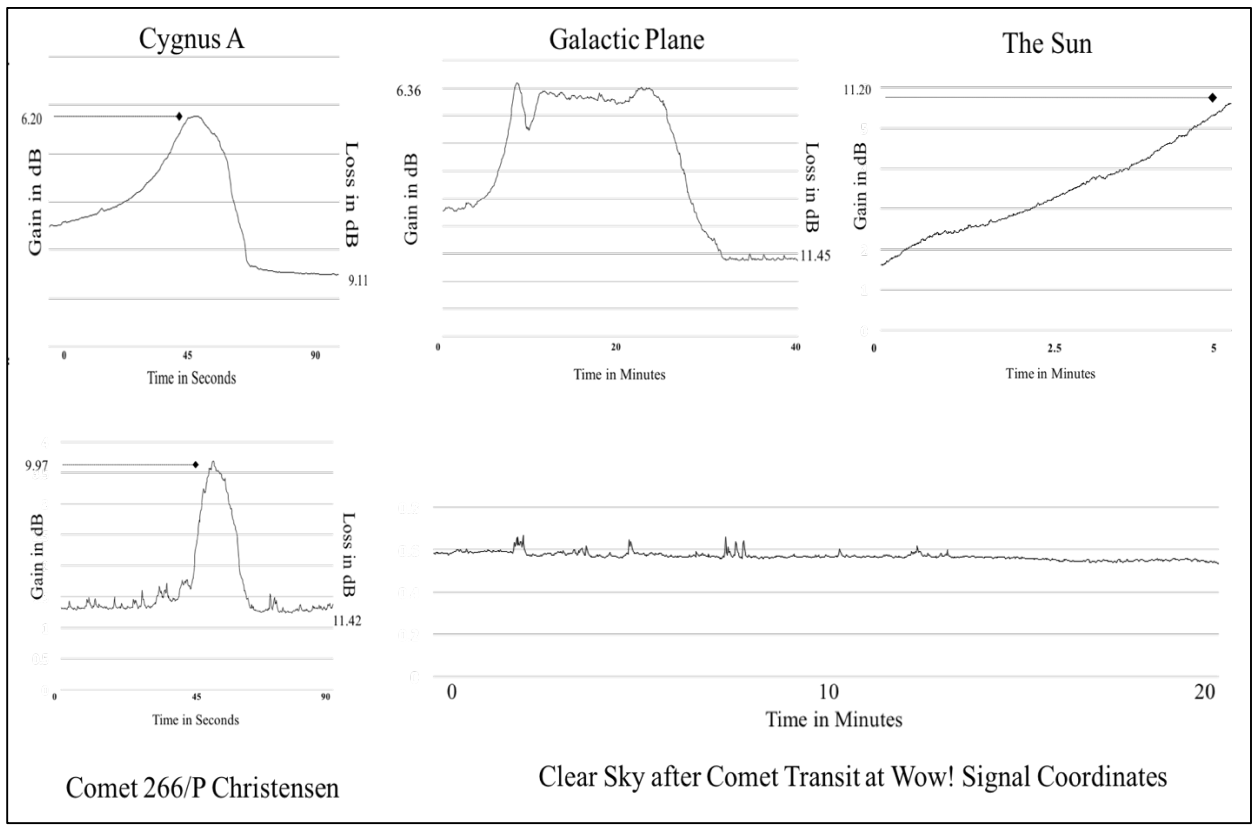

**Figure 23:** A Comparison of Neutral Hydrogen Observations

## Conclusions

In 2016, we proposed a hypothesis arguing that a comet and/or its hydrogen cloud was a strong candidate for the source of the "Wow!" Signal. From 27 November 2016 to 24 February 2017, we conducted 200 observations in the radio spectrum to validate the hypothesis. This investigation discovered that comets 266/P Christensen, P/2013 EW90 (Tenagra), P/2016 J1-A (PANSTARRS), and 237P/LINEAR emitted radio waves at 1420 MHz. In addition, the data collected during this investigation demonstrated there is a well-defined distinction between radio signals emitted from known celestial sources and comets, including comet 266/P Christensen.

We speculate that the strength of the original signal in 1977 would have been accounted for by the size of the Big Ear Radio Telescope (when compared with Site B) and/or the potential loss of mass from comet 266/P Christensen, which would have been considerably larger 40 years ago. In addition, while neutral hydrogen clouds have been observed around other comets (mostly from Lyman alpha spectra), determining the physical extent and density of the clouds around comets 266/P Christensen, P/2013 EW90 (Tenagra), P/2016 J1-A (PANSTARRS), and 237P/LINEAR were not the purposes of this investigation. To dismiss the source of the radio signal as emission from comet 266/P Christensen, we repositioned the telescope away from the comet and conducted clear sky observations when the comet was not near the coordinates of the "Wow!" Signal. During these clear sky observations, we detected no significant radio signal at 1420 MHz. This investigation, therefore, has concluded that cometary spectra are observable at 1420 MHz and that the 1977 "Wow!" Signal was a natural phenomenon from a Solar System body.

## About the Author

Prof. Antonio Paris is an Adjunct Professor of Astronomy at St. Petersburg College, FL and the Chief Scientist at the Center for Planetary Science. In 2016, in this journal, he published Hydrogen Clouds from Comets 266/P Christensen and P/2008 Y2 (Gibbs) are Candidates for the Source of the 1977 "Wow!" Signal—the hypothesis that led to this investigation. He is a member of the American Astronomical Society and is currently conducting research on gamma-ray pulsars, neutron stars, and black holes.

## Acknowledgments

This investigation could not have been completed without the hard work and dedication of Dennis Farr, Chief Technician at the Center for Planetary Science; Ryan Robertson, an undergraduate planetary science student at American Public University; Jeffrey Lichtman and Carl Lyster from Radio Astronomy Supplies; Edward Geiger, Consulting Telescope Engineer during this investigation, and Meade Instruments.

**References and Citations**

1. Shostak, S. Interstellar Signal from the 70s Continues to Puzzle Researchers. SETI (02 Dec. 2002). Accessed on 01 Oct. 2015. <http://archive.seti.org/epo/news/features/interstellar-signal-from-the-70s.php>.

2. Ehman, J. R. Wow! Signal—30th Anniversary Report. North American Astrophysical Observatory (28 May 2010). Accessed on 14 Oct. 2015. <http://www.bigear.org/Wow30th/wow30th.htm>.

3. The International Astronomical Union Minor Planet Center. Accessed on 21 Nov. 2015. <http://www.minorplanetcenter.net/>. Database: MPEC 2009-A03 P/2008 Y2 (Gibbs); MPEC 2008-U27 266P/Christensen.

4. SpectraCyber I/II™ 1420 MHz Hydrogen Line Spectrometer. Accessed on 27 Dec. 2016. <https://www.google.com/webhp?sourceid=chrome-instant&ion=1&espv=2&ie=UTF-8#q=spectracyber+pdf>.

5. SIMBAD Astronomical Database. Accessed on 01 Dec. 2016. <http://simbad.u-strasbg.fr/simbad/sim-basic?Ident=Cygnus+A&submit=SIMBAD+search>.

6. Baade, W.; Minkowski, R. (1954). "Identification of the Radio Sources in Cassiopeia (A), Cygnus A, and Puppis A". Astrophysical Journal. 119: 206. Accessed on 02 Dec. 2016. <http://articles.adsabs.harvard.edu/cgi-bin/nph-iarticle_query?1954ApJ...119..206B&data_type=PDF_HIGH&whole_paper=YES&type=PRINTER&filetype=.pdf>.

7. Dickey, J. M.; Lockman, F. J. (1990). "H I in the Galaxy". Annual Review of Astronomy and Astrophysics. 28: 215–259. Accessed on 12 Jan. 2017. <http://www.annualreviews.org/doi/10.1146/annurev.aa.28.090190.001243>.

8. Hydrogen Line Radio Astronomy, Moonbounce Communications and Radio Astronomy at K5SO. Accessed on 03 Jan 2017. <http://www.k5so.com/Radio_astronomy_HI_line.html>.